\documentstyle[epsfig]{jpsj}

\title{Spectral Properties of Three Dimensional Layered Quantum Hall Systems}
\author{Marcus {\sc Metzler}}
\inst{Department of Physics, Toho University, Miyama 2-2-1, Funabashi, 
Chiba 274-8510}
\recdate{}
\abst{
We investigate the spectral statistics of a network model for a three
dimensional layered quantum Hall system numerically. The scaling of
the quantity $J_0=\frac{1}{2}\langle s^2\rangle$ is used to determine
the critical exponent $\nu$ for several interlayer coupling strengths.
Furthermore, we determine the level spacing distribution $P(s)$ as
well as the spectral compressibility $\chi$ at criticality. We show
that the tail of $P(s)$ decays as $\exp(-\kappa s)$ with
$\kappa=1/(2\chi)$ and also numerically verify the equation
$\chi=(d-D_2)/(2d)$, where $D_2$ is the correlation dimension and $d=3$
the spatial dimension.} 
\kword{network model, layered system, level spacing distribution, spectral
compressibility, scaling, quantum Hall systems, multifractality}

\twocolumn
\begin{document}
\maketitle
Although the quantized Hall effect is usually attributed to
two-dimensional (2D) systems\cite{book} the possibility of an occurrence in
three-dimensional (3D) systems has been theoretically investigated early on
\cite{Azbel,AEW}. In addition to some precursory experimental findings
in 3D systems\cite{Mani,Zeitler} there have been two classes of quasi-3D
systems that show the integer quantum Hall effect: molecular
crystals\cite{Valfells} and multi-layer quantum wells formed by graded
GaAS/AlGaAs heterostructures\cite{Stoermer,WMMK}. Recently, the latter have
attracted significant theoretical interest. The existence of a
metallic phase for a finite range of energies and a new {\em chiral}
2D metallic phase on the surface have been proposed\cite{CD,Wang,BF}.
In ref.~\citen{CD} these 3D systems of 2D quantum Hall layers
coupled by weak interlayer tunneling have been investigated using a
network model. The existence of three phases: insulator, metal and
quantized Hall conductor accompanied by extended surface states
was shown by performing numerical transfer matrix calculations within
the network model.  

We will use the same kind of network model to investigate spectral
properties of the 3D layered system. In doing so we will apply the
methods used for the 2D Chalker-Coodington network
model\cite{Chalker_Coddington,LWK} which use the spectrum of the unitary
network operator $U$ to determine the statistics of the eigenvalue
spectrum of the underlying system\cite{Klesse_Metzler1,Metzler_Varga}. 
This method has the advantage that after determining the critical
energy $E_{\rm c}$ one can use the entire spectrum of eigenphases of
$U(E_{\rm c})$ to investigate the critical spectral statistics. 

The paper is structured in the following way:
After an introduction to the network model we will use finite size
scaling of the function $J_0(L,E)=\frac{1}{2}\langle s^2\rangle$, where $s$ is
the level spacing in units of the average level spacing $\Delta$, to
determine the critical point. At the critical point we are going to
investigate the level spacing distribution function $P(s)$ and the
number variance $\Sigma_2(N)=\langle(n-\langle n\rangle)^2\rangle$ of
an energy interval containing on average $N=\langle n\rangle$ levels.
We will see that, as in the 2D case, the following equation proposed in
ref.~\citen{CKL} holds:
\begin{equation}
  \label{eq:chi}
  \chi=\frac{\eta}{2d},
\end{equation}
where
$\chi=\lim_{N\rightarrow\infty}\lim_{L\rightarrow\infty}d\Sigma_2(N)/dN$ 
is the spectral compressibility, $d$ the spatial dimension of the system and
$\eta$ the diffusion exponent, which is related to the fractal
correlation dimension $D_2=d-\eta$\cite{Janssen_PhD}. We are also
going to show that the tail region of the level spacing distribution
shows an exponential decay at criticality, i.e. $P(s)\propto
\exp(-\kappa s)$ and that as proposed in ref.~\citen{AZKS}:
\begin{equation}
  \label{eq:kappa}
  \kappa=\frac{1}{2\chi}
\end{equation}

The network model for the 3D layered system developed by Chalker and
Dohmen (CD) is the natural generalization of the 2D Chalker-Coddington
(CC) network model\cite{CD,Chalker_Coddington}. It is a model for
electrons in 3D moving in a smooth disorder potential in
2D layers with a strong perpendicular magnetic field. In the
classical limit, reached when the correlation length $l_V$ of the
disorder potential $V$ is much larger than the magnetic length $l_{\rm c}$,
the motion of the electrons can be decoupled into three independent
components. The first with the smallest time scale is the cyclotron
motion about a guiding center. Secondly, we have oscillations parallel 
to the magnetic field on an intermediate time scale and for long times 
we have a drift motion along the equipotential contours of $V$ in the
plane perpendicular to the magnetic field. 
\begin{figure}[htbp]
  \begin{center}
   \epsfig{figure=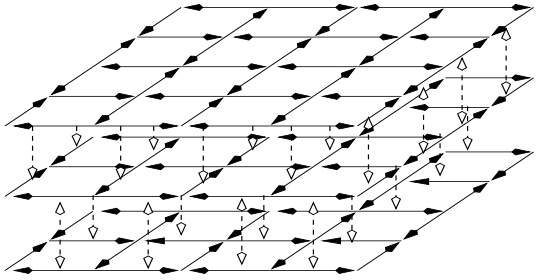,width=0.4\textwidth,height=0.2\textwidth}
    \caption{The 3D Chalker-Dohmen network. It consists of layers of
      the 2D Chalker-Coddington model coupled by interlayer scattering
      matrices. The dashed double arrows mark the interlayer coupling nodes.}
    \label{fig:cdn}
  \end{center}
\end{figure}

The network model simulates 
the system in the following way (see also Figure~\ref{fig:cdn}):
The unidirectional channels in the 2D planes represent the
equipotential contours of the potential in the plane. The random
length of the contour is simulated by a random phase added to the
electron at each channel link. 
The electrons travel along those lines until they reach a node of the
network which represents a saddle point of the potential. 
At the saddle point they are scattered. The scattering amplitude
$T=(1+\exp((E-u)/E_{\rm t})$ is
determined by the electrons energy $E$ and the saddle point energy
$u$, where $E_{\rm t}$ is the characteristic tunneling energy of the
potential\cite{Klesse_Metzler1,Fertig}. In the following all the
saddle point energies will be set to zero to avoid classical
percolation effects\cite{Klesse_Metzler1,Metzler_PhD,Metzler}.
In addition to the scattering at the nodes at each link a scattering
in the direction perpendicular to the plane takes place. This is
accomplished by a scattering matrix of the form 
\begin{equation}
  \label{eq:zscatter}
  S_z=\left( 
    \begin{array}{cc}
      \cos\phi&\sin\phi\\ -\sin\phi&\cos\phi
    \end{array} \right),
\end{equation}
giving us a transmission rate of $t=\sin\phi$, where $\phi$ is
the same for all interlayer scatterings.

A state or wave function of the network is defined by a normalized
vector $\Psi=(\psi_1, \dots,\psi_n)$ whose elements $\psi_i$ denote
the complex amplitudes on the $n=2L^3$ network channels, where $L$
is the system size. Its time evolution is determined by the
scattering matrices at the saddle points and the interlayer couplings
which can be expressed by one unitary matrix $U$ the so called network
operator:
\begin{equation}
  \label{eq:U}
  \Psi(\tau+1)=U\Psi(\tau).
\end{equation}
Actually, $U$ can be written as the product of a matrix $U_{2d}$ built 
from the independent network operators of the respective layer systems 
and an interlayer scattering matrix $S$ which takes care of the
coupling of those systems, i.e. $U=U_{2d}S$. Both matrices are sparse
with only 2 non-zero elements per column, which means that $U$ is also sparse
with 4 non-zero elements per column. U is energy dependent through the 
energy dependence of the scattering matrices in the layers.

An energy eigenstate $\Psi_E$ of the underlying electronic system is a 
solution of the stationarity equation:
  \begin{equation}
    \label{stat}
    U(E)\Psi_E=\Psi_E
  \end{equation}
which corresponds to an eigenstate of $U(E)$ with eigenvalue unity. Such
an eigenstate will occur only at discrete values of $E=E_{\rm c}$ which form 
the energy spectrum of the modeled system. However, it has been shown
that the eigenphases $\omega_\alpha(E)$ of the unitary network operator
$U(E)$ determined by the equation
\begin{equation}
  \label{eq:Ueigen}
  U(E)\Psi_\alpha=e^{i\omega_\alpha(E)}\Psi_\alpha
\end{equation}
show the same statistics as eigenenergies $E_n$ close to
$E$\cite{Klesse_Metzler2}. Since the arguments in
ref.~\citen{Klesse_Metzler2}, given for the Chalker-Coddington model,
hold for any network model that includes an energy parameter we can
use the so called quasienergies $\omega_\alpha(E)$, which can be
considered an excitation spectrum at energy $E$, to determine the
spectral properties of the 3D system.

Since the 3D network model exhibits three different phases, i.e. two
phase transitions, our first task is to determine the critical points
of those transitions. It has already been established in
ref.~\citen{CD} that at energy $E_{\rm c_1}(t)(<0)$ the network exhibits an
insulator-metal (IM) transition and symmetrical to the $E=0$ axis at
energy $E_{\rm c_2}(t)=-E_{\rm c_1}(t)$ a metal-quantum Hall conductor (MQH)
transition for every $t$.
The states in the quantum Hall conductor are all localized as in the
insulator phase except for the fact that for hard wall boundary
conditions there exist extended surface states separated from the
localized bulk states. In our investigation we are using mostly
periodic boundary conditions, but we will comment on the hard walled
systems if results differ significantly.

The width $W=E_{\rm c_2}-E_{\rm c_1}$ of the metallic phase was shown to depend
on the coupling $t$ in the following way: $W(t)\propto
t^{1/\nu_{2d}}$, where $\nu_{2d}\approx 2.3$ is the critical exponent
of the localization length in 2D quantum Hall Systems\cite{book}. 
This knowledge was used to narrow the range for our search of the
critical energies. 
\begin{figure}[htbp]
  \begin{center}
        \epsfig{figure=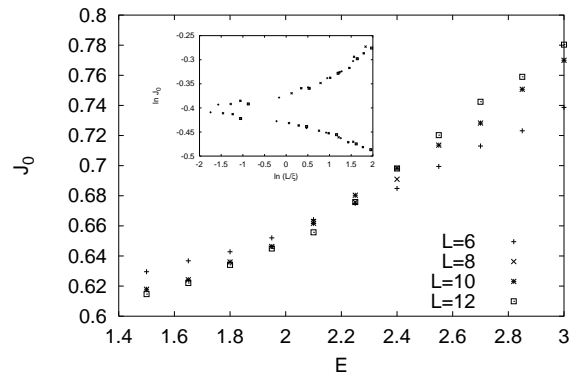,angle=270,width=0.43\textwidth}
    \caption{The scaling variable $J_0$ as a function of $E$ for
      different system sizes and $t=0.2$. At $E\approx 2.2$ all the
      curves meet in a single point. The inset shows the one-parameter
      dependence of $J_0$ on $L/\xi(E)$.}
    \label{fig:t0.2pos}
  \end{center}
\end{figure}

The critical energies are determined by finite size scaling.
As a scaling variable the quantity $J_0=\frac{1}{2}\langle s^2 \rangle$ is
used. This function reflects the transition of the spectral
correlations from Gaussian unitary ensemble (GUE) to Poisson statistics when we
cross from the metallic phase into a localized regime. The values
$J_0=0.590$ in the metallic and $J_0=1$ in the localized regime are
well known from random matrix theory (RMT)\cite{rmt}.
\begin{figure}[htbp]
  \begin{center}
        \epsfig{figure=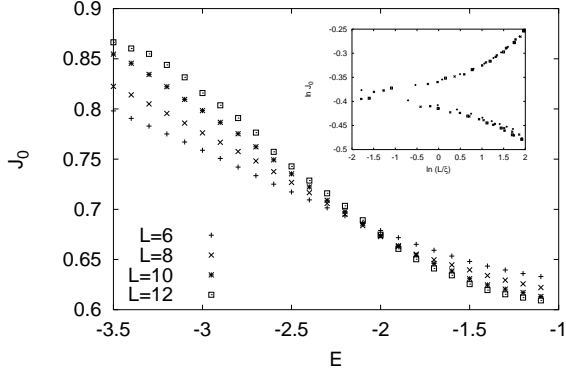,angle=270,width=0.43\textwidth}
    \caption{The scaling variable $J_0$ as a function of $E$ for
      different system sizes and $t=0.2$. At $E\approx -2.2$ all the
      curves meet in a single point. The inset shows the one-parameter
      dependence of $J_0$ on $L/\xi(E)$.}
    \label{fig:t0.2neg}
  \end{center}
\end{figure}

In Figs.~\ref{fig:t0.2pos} and \ref{fig:t0.2neg} we can see
the function $J_0(E)$ at $t=0.2$ for different system sizes $L$ and energies
$E>0$ and $E<0$, respectively. The curves for different $L$ in both
figures intersect in single points at $E_{c_1}\approx -2.2$ and
$E_{c_2}\approx 2.2$ with $J_0^{\rm c}\approx 0.68$ in both cases. These
points mark the IM and MQH transitions. Assuming the validity of
one-parameter scaling, i.e. $J_0(E,L)=f(L/\xi(E))$, and the existence of 
the localization length $\xi(E)$ that shows the power law behavior
$\xi(E)\propto |E-E_{\rm c}|^{-\nu}$, one can re-scale $L$ so that the data of
$J_0$ as a function of $L/\xi$ for consecutive energies $E$ overlap
with each other. Within statistical accuracy all points fall onto one
common curve independent of $L$ and $E$. This can be seen in the
insets of the two figures where $J_0(E,L)$ is plotted against $L/\xi(E)$. 
\begin{table}[h]
  \begin{center}
    \begin{tabular}{@{\hspace{\tabcolsep}\extracolsep{\fill}}rrrrrrr}
      \hline
      $t$&$E_{\rm c}$&&$\nu$&&$J_0^{\rm c}$&\\
      \hline
      $0.1$&$-1.6 $&$\pm 0.2 $&$1.2 $&$\pm 0.5 $&$0.73 $&$\pm 0.05$\\
      $0.2$&$ 2.17$&$\pm 0.03$&$1.33$&$\pm 0.18$&$0.669$&$\pm 0.003$\\
      $0.2$&$-2.04$&$\pm 0.02$&$1.27$&$\pm 0.08$&$0.682$&$\pm 0.002$\\
      $0.4$&$ 3.08$&$\pm 0.03$&$1.5 $&$\pm 0.2 $&$0.68 $&$\pm 0.01$\\
      $0.5$&$ 3.33$&$\pm 0.03$&$1.45$&$\pm 0.1 $&$0.69 $&$\pm 0.01$\\
      \hline
    \end{tabular}
    \caption{The results for the 4-parameter fit of eq.~(\ref{eq:expand}).}
    \label{tab:4pf}
  \end{center}
\end{table}
In order to calculate the critical exponent $\nu$ we expand the
finite-size scaling function around the critical point:
\begin{equation}
  \label{eq:expand}
  J_0(E,L)\approx J_0^{\rm c}+B(E-E_{\rm c})L^{1/\nu}.
\end{equation}
We can now either determine $\nu$, $J_0^{\rm c}$, $E_{\rm c}$ and $B$ by a
four-parameter fit or we can take the energy derivative of
eq.~(\ref{eq:expand}), which gives us $dJ_0(E,L)/dE\propto L^{1/\nu}$
and perform a two parameter linear fit to a double-log plot of the
slope of $J(E)$ as a function of $L$ at the critical point $E_{\rm c}$.
The results for the 4-parameter fit are shown in Table~\ref{tab:4pf}
for different values of the coupling $t$. 
The linear fit of the slope data yields essentially the same results
as the 4-parameter fit (see Table~\ref{tab:linfit}).
The results for the exponent $\nu$ are in agreement with earlier
numerical results for layered 3D systems\cite{OKO,CD} ($\nu=1.35\pm
0.15$ and $\nu=1.45\pm 0.25$) and 3D Anderson models\cite{SO,HS1,HS2}
($\nu=1.43\pm 0.06$ and $\nu=1.34\pm 0.10$) with broken time reversal symmetry.
The values for the critical energy also agree with the results of Chalker and
Dohmen\cite{CD}. 
\begin{table}[h]
  \begin{center}
    \begin{tabular}{@{\hspace{\tabcolsep}\extracolsep{\fill}}rrrrr}
      \hline
      $t$&$E_{\rm c}$&&$\nu$&\\
      \hline
      $0.1$&$-1.6$&$\pm 0.05$ &$1.2 $&$\pm 0.2 $ \\
      $0.2$&$-2.2$&$\pm 0.05$ &$1.34$&$\pm 0.06$ \\
      $0.2$&$ 2.2$&$\pm 0.1 $ &$1.32$&$\pm 0.04$ \\
      $0.4$&$ 3.1$&$\pm 0.1 $ &$1.6 $&$\pm 0.4 $ \\
      $0.5$&$ 3.4$&$\pm 0.1 $ &$1.3 $&$\pm 0.1 $ \\
      \hline
    \end{tabular}
    \caption{The results for $\nu$ obtained from the derivatives
      $dJ_0/dE$ at $E_{\rm c}$.}
    \label{tab:linfit}
  \end{center}
\end{table}

\begin{figure}[htbp]
  \begin{center}
        \epsfig{figure=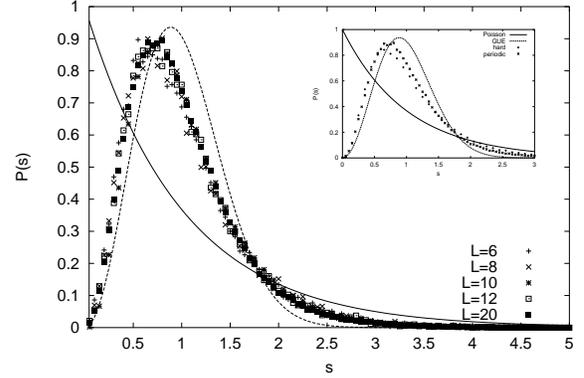,angle=270,width=0.43\textwidth}
    \caption{The critical level spacing distribution $P(s)$ for
      $t=0.2$ and several system sizes compared to the Poisson and
      GUE distributions. The inset shows the critical $P(s)$
      for two different boundary conditions.}
    \label{fig:pos}
  \end{center}
\end{figure}

Now that we know the critical energies $E_{\rm c}(t)$ we can concentrate on the
investigation of the spectral properties at criticality.
We start by looking at the critical level spacing distribution $P(s)$.
We are going to compare it to the Poisson distribution
($P_{\rm P(s)}=\exp(-s)$) in the localized and the GUE distribution
($P_{\rm GUE}(s)=\frac{32}{\pi^2}s^2\exp\left( -\frac{4}{\pi}s^2\right)$)
in the metallic regime known from RMT\cite{rmt}. The critical $P(s)$
shows characteristics of both of these distributions. In
Fig.~\ref{fig:pos} we can see the critical $P(s)$ for $t=0.2$ and
several system sizes. As expected they are independent of the system
size and all fall on the same curve. They also obviously differ strongly from
the Poisson and GUE distributions. Those distributions were obtained
from networks with periodic boundary conditions. The inset of
Fig.~\ref{fig:pos} shows a comparison of the periodic boundary case with a
critical distribution of a system with hard wall boundaries (HWB) in one
direction. We can see that in the HWB case $P(s)$ is shifted towards
the Poisson distribution which means a different $J_0^c\approx 0.7$. 
This has also been observed in other 3D systems\cite{BMP}. We also found a 
slightly increased value for $E_c$ which if confirmed has to be investigated 
in later publications (see also refs.~\citen{EM} and \citen{PS}).

\begin{figure}[htbp]
   \begin{center}
        \epsfig{figure=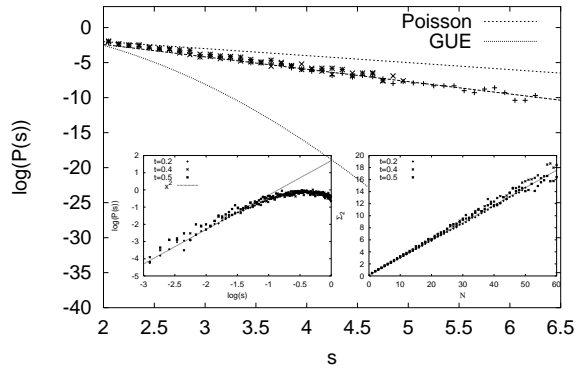,angle=270,width=0.43\textwidth}
    \caption{The tail of the critical $P(s)$ for $t=0.2,0.4,0.5$
      compared to the GUE and Poisson case. The left inset shows the
      small $s$ behavior of $P(s)$ compared to the function $s^2$. The
      right inset is the level number variance at criticality.} 
    \label{fig:shape}
  \end{center}
\end{figure}

A closer investigation of the shape of the critical distribution
reveals that for small values of $s$ it resembles the GUE case. The
inset in Fig.~\ref{fig:shape} shows the small $s$ behavior and we can see
that $P(s)\propto s^2$ for all couplings $t$. This reflects the level
repulsion due to the multifractal eigenstates which, like 
the homogeneously smeared out metallic eigenstates, cover the entire
system. On the other hand, a look at the behavior for large values of
$s$ shows an exponential decay of the tail of $P(s)$ (see
Fig.~\ref{fig:shape}) which is typical for the Poisson distribution and in 
contrast to the Gaussian tail of the GUE case.
\marginpar{\hfill\fbox{Fig.5}}
Compared to the Poisson case, the critical distribution decays
faster with $P(s)\propto \exp(-\kappa s)$ where $\kappa=1.76\pm 0.04$
for $t=0.2$, which is the coupling where we got the most statistics.
According to ref.~\citen{AZKS}, which predicted this exponential decay,
the factor $\kappa$ is connected to the spectral compressibility
$\chi$ via eq.~(\ref{eq:kappa}). This leads us to the next step in our
investigation, the determination of $\chi$.

In the right inset of fig.~\ref{fig:shape} we can see the level number
variance $\Sigma_2$ whose slope gives us $\chi$. The results for 
the different coupling strengths all agree with each other and we
again took the value for $\chi$ from our data for $t=0.2$. We
obtained: $\chi=0.286\pm 0.003$ which gives us according to
eq.~(\ref{eq:kappa}) $\kappa=1.75\pm 0.01$ which is in good agreement
with the directly obtained value of $\kappa$. These values for $\chi$
and $\kappa$ are also quite close to the ones obtained in
ref.~\citen{ZK} for a 3D Anderson model with time reversal symmetry.
Furthermore, we can connect the spectral compressibility directly to
the multifractal critical wave functions. This was done in
ref.~\citen{CKL} resulting in eq.~(\ref{eq:chi}). If we use $\chi$ we
obtain a value of $\eta=1.72\pm 0.02$ or $D_2=1.28\pm 0.02$. If we
compare $D_2$ with direct results for the network model ($1.3\pm
0.1$\cite{HK}) or 3D Anderson models without time reversal symmetry
($1.7\pm 0.2$\cite{OK},$\approx 1.5$\cite{Terao}) we see that
eq.~(\ref{eq:chi}) holds within the errors.
Although the result for $D_2$ taken from eq.~\ref{eq:chi}, which fits very
well to the direct result from ref.~\citen{HK}, is a little smaller
than those of other investigations.

In conclusion, the numerical investigation of the spectral statistics
of a network model for a three dimensional layered quantum Hall system for
different coupling strengths between the layers showed that the
phase transitions from metallic to localized regimes all have the same 
characteristics. The critical exponent $\nu\approx 1.3$ is in
agreement with other results for similar systems.
Furthermore, the relations (\ref{eq:chi}) and (\ref{eq:kappa}) were
verified to a high degree of accuracy within the network model.  
This is an improvement to other 3D models (all of them with time reversal
symmetry) where the two equations could also 
be verified but with much larger errors\cite{CKL}.
This shows that the usage of network models, apart from their
advantages for the investigation of time evolution, gives us a more efficient
way to investigate spectral properties.

The author would like to thank Y. Ono for valuable discussions and the Deutsche
Forschungsgemeinschaft (DFG) for their financial support.


\end{document}